\documentclass[universe,article,accept,pdftex,moreauthors]{Definitions/mdpi} 
\graphicspath{{figs/}}
\usepackage[T1]{fontenc}
\usepackage{xcolor}
\usepackage{amsmath}
\usepackage{graphicx}
\newcommand{\eb}{\begin{equation}}
\newcommand{\ee}{\end{equation}}

\definecolor{rkka}{RGB}{219,66,32}

\firstpage{1} 
\pubvolume{9}
\issuenum{11}
\articlenumber{463}
\pubyear{2023}
\copyrightyear{2023}
\externaleditor{Academic Editor: Giacomo Tommei}
\datereceived{6 October 2023 } 
\daterevised{20 October 2023 } 
\dateaccepted{23 October 2023 } 
\datepublished{28 October 2023 } 
\hreflink{https://doi.org/10.3390/\linebreak universe9110463} 

\Title{Secular Orbital Dynamics of the Possibly Habitable Planet K2-18 b with and without the Proposed Inner Companion}

\TitleCitation{Secular Orbital Dynamics of the Possibly Habitable Planet K2-18 b with and without the Proposed Inner Companion}

\Author{{Valeri} 
 V. Makarov $^{1,}$*\orcidA{} and Alexey Goldin $^{2}$}

\AuthorNames{Valeri V. Makarov and Alexey Goldin}

\AuthorCitation{{Makarov,} 
 V.V.; Goldin, A.}

\address{%
$^{1}$ \quad United States Naval Observatory, 3450 Massachusetts Ave. NW, Washington, DC {20392-5420,} 
 USA\\
$^{2}$ \quad Teza Technology, 150 N Michigan Ave, Chicago, IL {60601}, USA; alexey.goldin@gmail.com}

\corres{Correspondence: valeri.makarov@gmail.com}

\abstract{The transiting planet K2-18 b is one of the best candidates for a relatively nearby world harboring biological life.
The long-term orbital evolution of this planet is investigated using theoretical and purely numerical techniques for two possible configurations:
a single planet orbiting the host star, and a two-planet system including the proposed inner planet close to the 4:1 mean motion rationalization. The emphasis
is made on the secular changes of eccentricity and orbital inclination, which are important for the climate stability of the planet. It is demonstrated
that the secular orbital dynamics of planet K2-18 b with an internal companion are accurately represented by the periodic eccentricity and inclination
exchange on the time scales of a few Kyr. A single planet is not expected to experience fast orbital changes, with the much weaker tidal and rotation-driven
perturbations mostly reflecting in a slow periastron and nodal precession. The tidal decay of the orbit is too insignificant on the time scale of the stellar age. However, the conditions for the habitability of a single  K2-18 b planet are much improved if, like the Earth, it rotates faster than the mean motion and its rotation angle is tilted by a hypothetical moon. Milankovi{\'c}'s cycles of the habitable planet's climate are discussed for both configurations.}


\begin{document}

\section{Introduction}
K2-18 is an M2.8 main sequence dwarf at a distance of 38.1 pc from the Sun. It harbors at least one planet (K2-18 b) detected in the photometric light curves of the extended Kepler mission \citep{2015ApJ...806..215F, 2015ApJ...809...25M}. The orbital period of this planet is 32.94 d \citep{2018AJ....155..257S}, and its estimated bulk density of 3300 kg m$^{-3}$ suggests a terrestrial central body covered with a massive ocean~\citep{2017A&A...608A..35C}. Observing the planetary transits with the Hubble Space Telescope, ref.~\citet{2019ApJ...887L..14B}, the presence of water vapor and clouds in the atmosphere of planet b were detected. Coupled with the estimated insolation flux, which is only slightly higher than the insolation of Earth~\citep{2017ApJ...834..187B}, these properties make K2-18 b one of the best candidate habitable worlds with thriving biological life. Precision radial velocity observations confirmed the presence of planet b, but also indicated a possible inner companion with a period of $8.962\pm0.008$ d \citep{2017A&A...608A..35C}. The mass and radius of this proposed planet c are only slightly smaller than the values for planet b. A downward revision of the projected mass $M_c\sin i$ followed a more detailed analysis of the available radial velocity data from two separate instruments \citep{2019A&A...621A..49C}. A novel technique to process combined spectroscopic data further strengthened the case for an inner planet and updated the estimated mass to $6.99\,M_{\bigoplus}$ and the period to 9.2 d \citep{2022MNRAS.517.5050R}. Still, doubts linger in the reality of the inner planet, and it is listed as a ``controversial'' entry in the NASA Exoplanet Archive {available online (}
{\url{https://exoplanetarchive.ipac.caltech.edu}, {accessed on 27 August 2023}
). The tentative inner planet is not transiting; thus, its orbital inclination should differ from that of planet b by at least 2$^\circ$. The period ratio is approximately 3.7, which is close to the potential 4:1 mean motion resonance (MMR). Are such systems stable long-term? 

\textls[-15]{Another consideration to be taken into account is the rate of rotation of the host star. The projected surface velocity from the broadening of the spectral lines is \mbox{$v\,\sin i=2.0$ km s$^{-1}$}~\citep{2018A&A...612A..49R},} which yields an upper bound rotation period of 8.8 d for the estimated radius. This is close to the orbital period of the tentative planet c. Photospheric spots and plages can cause measurable modulation of the radial velocity curve, which, in most cases, manifests itself as an enhanced jitter because of the short life-times of these formations and their small effective area \citep{2009ApJ...707L..73M}. Magnetically active stars, on the other hand, are known to have persistent structures (such as spot groups) that can last for multiple rotations and occupy a sizable portion of the visible disk. They cause quasi-sinusoidal variations of the Doppler shift, which can be confused with an exoplanet signal. K2-18, however, does not show significant signs of magnetic activity or binarity in the available data. Persistent photometric structures are not expected for this star. Precision radial velocity measurements over extended timescales should resolve this open issue by confirming or rejecting the persistent phase of the radial velocity modulation.

The goal of this study is to investigate the consequences of both options, the single planet and two-planet system, for the secular orbital evolution of
the possibly habitable planet b. Variations of the semimajor axis, eccentricity, and inclination are very important factors for the considerations of habitability and climate modeling. The plan of this paper is as follows. In Section \ref{ex.sec}, we investigate the basic secular exchange of eccentricity and inclination for the two-planet option. Our approach combines high-accuracy and long-term numerical simulations with the theoretical models derived from the classical Laplace--Lagrange analysis.

\section{Eccentricity and Inclination Exchange in a Two-Planet K2-18 System}\label{ex.sec}

In this section, we investigate the fundamental process of the orbital momentum exchange between two planets and its manifestations in the secular orbital evolution of
planet \mbox{K2-18 b} if the proposed inner planet c is present. A series of numerical simulations was performed using the symplectic integrator WHFast \citep{2015MNRAS.452..376R}, which is part of the REBOUND package (\url{https://rebound.readthedocs.io/en/latest/}, {accessed on 19 October 2023}). This code provides high-speed computation in conservative dynamical systems. Most of the computations were made for $3\times 10^5$ yr with 50 steps per period of the inner planet.
The basic characteristics of the planet system adopted in our integration are listed in Table \ref{param.tab}. The orbital inclination and eccentricity of the two orbits are practically unknown. Specifically, only a very wide upper limit on eccentricity can be derived from the available transit and radial velocity data, which is not helpful for the accurate characterization of orbital dynamics. It is logical to assume that both planets' eccentricities are ``small", while the mutual inclination is larger than a few degrees (from the fact that the inner planet is not transiting). We, therefore, investigate a grid of possible initial configurations. The initial eccentricity of the inner planet is always \mbox{$e_1(0)=0.01$,} and the accepted values for the outer planet are $e_2(0)=0.01,\,0.05,\, 0.1,\,0.2$. Each of these configurations are also integrated on a grid of initial inclinations \mbox{$i_2(0)=5,\,10,\,20,\,40^\circ$,} thus making a set of 16 different configurations. The initial orientation of the inner orbit is set to $i_1(0)=0$, $\omega_1(0)=0$, and $\Omega(0)=0$ by the choice of the fixed non-rotating coordinate frame. We also set $\omega_2(0)=0$ and $\Omega_2(0)=0$. These Euler angles  are found to be rapidly varied on the time scale of $\sim$100~yr, and no loss of generality is expected from this choice. 

Secular gravitational interactions between the planets and the star cause the orbital elements to vary with time. In the asymptotic limit of small eccentricity and mutual inclination, this behavior is predicted by the classical Laplace--Lagrange linear perturbation theory, which was initially developed for the Solar system dynamics, and specifically, for the Jupiter--Saturn pair \citep{1950USNAO..13...81B,1999ssd..book.....M}. Under these conditions, the corresponding model is
\begin{eqnarray}
e_j(t)&=&\sqrt{A_1+A_2\,\cos(2\pi t/p_e+\phi_e+\pi\,j)},\\
i_j(t)&=&\sqrt{B_1+B_2\,\cos(2\pi t/p_i+\phi_i+\pi\,j)},
\label{model2}
\end{eqnarray}
where $j=1$ designates the inner planet, and $j=2$ is for the outer planet. We note that in the special case $e_1(0)=0$, the equations for $e_j(t)$ become equivalent to the mirrored sine model with $A_1=A_2=A_e^2/2$ and  $\phi_e=0$, which was used in ref. ({Makarov et al., 2023, submitted}
). The phase shift $\pi\,j$ assures that the oscillations of both eccentricities (and both inclinations) are in the opposite direction, so that when one planet reaches its maximum eccentricity, the other planet reaches the minimum value. When both initial eccentricities are nonzero, $A_1\neq A_2$, and the curve looks like a sinusoid with sharper minima. The periods of oscillation $p_e$ and $p_i$ are equal for both planets. The common phase for the tested configurations is either 0 or $\pi$.

Equation (\ref{model2}) provides excellent fits to the osculating orbital elements for all trial initial eccentricity values and for initial inclinations $i_2(0)$ up to $20^\circ$. An example of the output simulated eccentricities and inclinations for both planets integrated over 30 Kyr is shown in Figure \ref{per.fig}. The initial conditions are: $e_1(0)=0.01$, $e_2(0)=0.05$, $i_1(0)=0$, $i_2(0)=5^\circ$, $\omega_1(0)=\omega_2(0)=0$, and $\Omega_1(0)=\Omega_2(0)=0$, and only the first 10 Kyr of the data are shown with black dots sampled every 10 yr. The fits (shown with red curves) follow the output data so closely that the black dots are barely seen. The fit for $i_1(t)$, with its minima close to the initial $i_1(0)=0$, degenerates to
$i_1(t)=\sqrt{2B_1}|\sin(4\pi t/p_i)|$, as explained above. The full amplitudes of eccentricity and inclination variations are significantly smaller for planet 2 than for planet 1. This is expected from the total angular momentum preservation, and is true for all configurations. The ``observed minus calculated'' residuals are smaller than the oscillation amplitudes by two orders of magnitude in this example, but they reveal the presence of additional eigenfrequencies that are not captured by the model. These additional eigenfrequencies, however, are sensitive to the initial mutual inclination of the orbits, and become obvious at $i_2(0)=20^\circ$.

\begin{figure}[H]
\begin{adjustwidth}{-\extralength}{0cm}
\centering 

\begin{tabular}{cc}
\includegraphics[width=.55\textwidth]{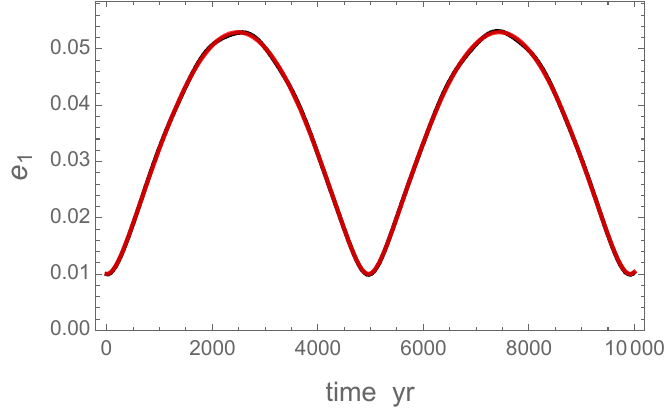}&
\includegraphics[width=.53\textwidth]{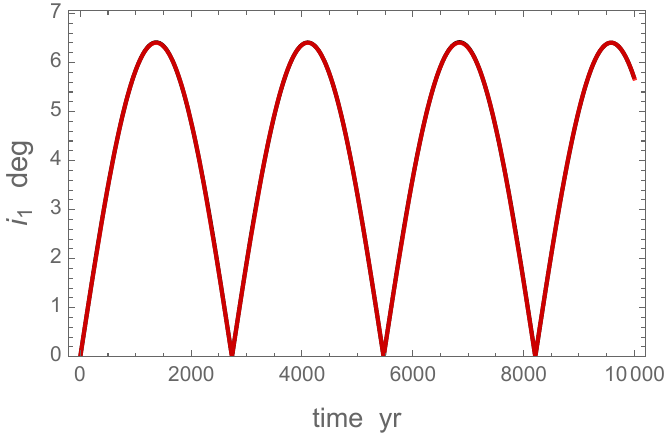}\\
\includegraphics[width=.55\textwidth]{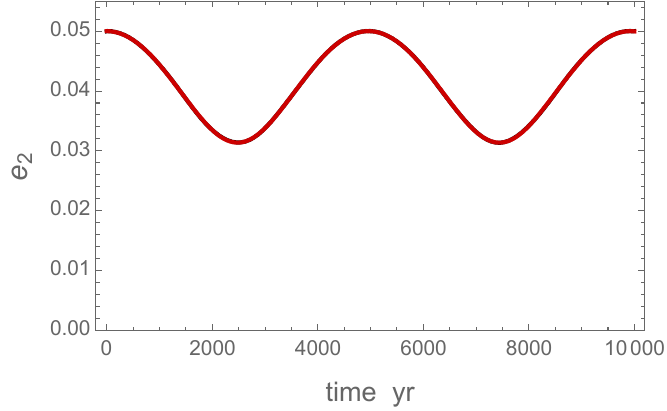}&
\includegraphics[width=.53\textwidth]{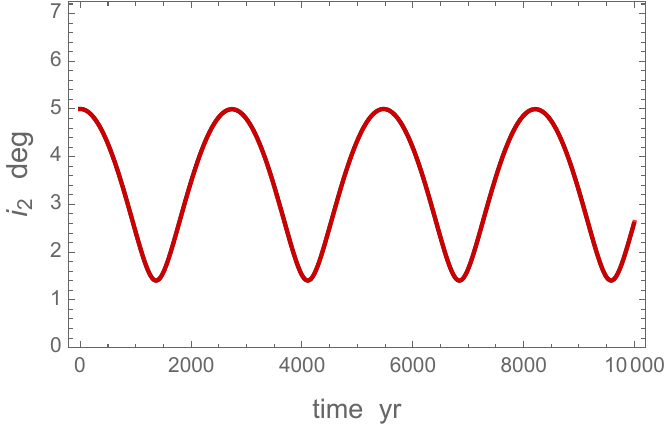}
\end{tabular}
\end{adjustwidth}\vspace{-6pt}
\caption{{Secular} 
 evolution of K2-18 planets in a two-planet configuration. Numerical integration of orbital motion was performed for 30 Kyr with
the initial conditions: $e_1(0)=0.01$, $e_2(0)=0.05$, $i_1(0)=0$, $i_2(0)=5^\circ$, $\omega_1(0)=\omega_2(0)=0$, and $\Omega_1(0)=\Omega_2(0)=0$. 
The black curves show the actual output of the numerical simulations. The red curves show the best fits with the theoretical model~(\ref{model2}). \textbf{Upper left} panel:
inner planet K2-18 c eccentricity. \textbf{Upper right} panel: inner planet K2-18 c inclination in degrees.  \textbf{Lower left} panel:
outer planet K2-18 b eccentricity. \textbf{Lower right} panel: outer planet K2-18 b inclination in~degrees. \label{per.fig}}

\end{figure}

\begin{table}[H]
\caption{{Assumed} 
 and adopted physical parameters of a two-planet K2-18 system.\label{param.tab}
}
\begin{tabularx}{\textwidth}{CCCCC}
\toprule
\textbf{Object} & \textbf{Mass} & \textbf{Period, d} & \textbf{Radius} \\
\midrule
K2-18 & 0.36 $M_{sun}$ &   &  0.46 $R_{sun}$   \\
K2-18 b & 0.0272 $M_{jup}$ & 32.94  &  0.233 $R_{jup}$  \\
K2-18 c & 0.0236 $M_{jup}$ & 8.962  &  0.22 $R_{jup}$  \\
\bottomrule

\end{tabularx}
\end{table}

The emerging picture is consistent with previous finds from long-term integrations of other exoplanet systems with period ratios closer to the 2:1 MMR. The main-mode periods $p_e$ and $p_i$ are weakly dependent on the initial $e$ and $i$, as long as they are close to zero. The general dependence of $p_i$ on $i_2(0)$, however, is exponential. For the configuration in \mbox{Figure \ref{per.fig},} for example, the fitted periods are $p_e=4960$ yr and $p_i=2737$ yr. With the same initial conditions, but with $i_2(0)$ set to $20^\circ$ instead of $5^\circ$, the periods become \mbox{9050 yr} and 3206 yr, respectively. Furthermore, if the initial inclination is set to $40^\circ$, model (\ref{model2}) is no longer valid for eccentricity, as shown in Figure \ref{e40.fig}. Both planets' eccentricities vary non-periodically in a wide range. For the inclination of the outer planet, model (\ref{model2}) still provides an adequate, but imperfect, fit (right panel). The estimated period $p_i$ is 4722 yr.

\vspace{3pt}
\begin{figure}[H]

\begin{adjustwidth}{-\extralength}{0cm}
\centering 

\includegraphics[width=.62\textwidth]{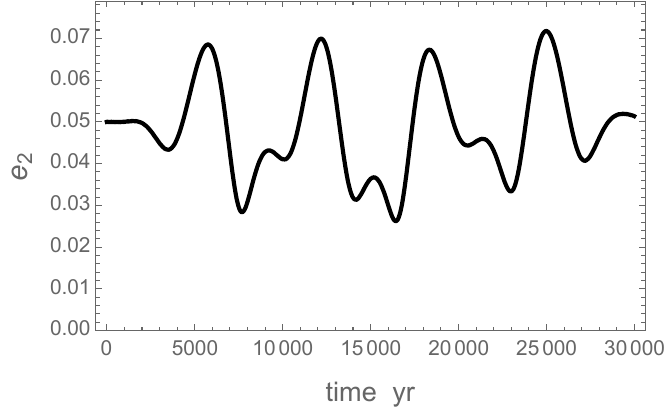}
\includegraphics[width=.62\textwidth]{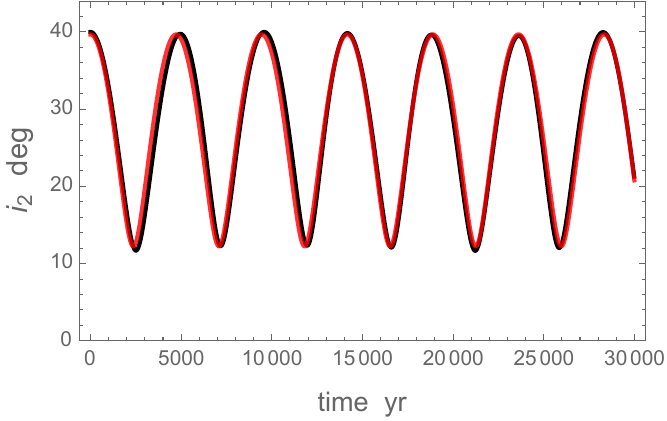}
\end{adjustwidth}
\caption{{Secular} 
 evolution of K2-18 planets in a two-planet configuration. The numerical integration of orbital motion was performed for 30 Kyr with
the initial conditions: $e_1(0)=0.01$, $e_2(0)=0.05$, $i_1(0)=0$, $i_2(0)=40^\circ$, $\omega_1(0)=\omega_2(0)=0$, and $\Omega_1(0)=\Omega_2(0)=0$. 
The black curves show the actual output of the numerical simulations. The red curve show the best fits with the theoretical model (\ref{model2}). \textbf{Left} panel:
outer planet K2-18 b eccentricity. \textbf{Right} panel: outer planet K2-18 b inclination in degrees.  \label{e40.fig}}
\end{figure}

\vspace{-6pt}

\section{Apsidal Precession and Nodal Recession in a Two-Planet K2-18 System}
\label{prec.sec}
We are not aware of an accurate and tested analytical theory describing the secular and long-period evolution of the node and periapse position in the three-body non-restricted
problem with nonzero eccentricities and inclinations. In some papers, e.g., \citep{2002ApJ...564.1019M}, the impact of the inner planet is approximated with an additional quadrupole moment $J_2$ assigned to the star, and the well-developed Lagrange theory for multipole perturbations is then employed. Essentially, the inner planet is replaced with a stationary gravitating ring of a uniform density with a radius of $a_1$. Our numerical experiment for K2-18 with two planets reveals that this approximation gives an incomplete and distorted picture of the complex behavior of the two orbits. Specifically, it predicts that the line of apsides precesses for the outer planet (if mutual inclination is less than $\arcsin(2/\sqrt 5)$), and the line of nodes recesses. The likely reason for this inadequacy is that the quadrupole approximation does not capture the exchange of angular momentum between the two planets.

From our benchmark simulation with the initial parameters $e_1(0)=0.01$, $e_2(0)=0.05$, $i_1(0)=0$, $i_2(0)=5^\circ$, $\omega_1(0)=\omega_2(0)=0$, and $\Omega_1(0)=\Omega_2(0)=0$,
we find that the inner planet's $\omega_1(t)$ shows both types of secular perturbation: a linear prograde drift (precession) and a long-period variation. The rate of precession can be expressed as 
\eb 
\omega_1(t)=2\pi\,t/P_{\omega 1},
\ee 
where $P_{\omega 1}$ is the period of periapse circulation. For the tentative inner planet in this configuration, we estimate $P_{\omega 1}=3790$ yr, which is significantly shorter than the eccentricity exchange period $p_e=4960$ yr. Overlaid on this drift, there is a long-period (hereafter, LP) oscillation with a period of about 5000 yr and a full amplitude of $\sim$1.5 rad. The oscillation is markedly non-sinusoidal. Remarkably, the $\Omega_1(t)$ curve does not include any LP variations but shows a steady linear recession with a circulation period $P_{\Omega 1}=5500$ yr. Perhaps, the results for the outer planet 2 are more important. The $\omega_2(t)$ curve includes both the linear prograde drift and LP components. The estimated periapse circulation period is $P_{\omega 1}=13,000$ yr, which is much slower than the inner planet circulation. Figure~\ref{omegas.fig}, on the left panel, shows the LP component of $\omega_2(t)$ after the subtraction of the linear precession term. The variations are vaguely periodic but non-sinusoidal, with uneven extrema and faster declining slopes than the inclining slopes. Remarkably, the line of nodes' temporal behavior depicted in the right panel without any manipulation with the output data displays only a
non-sinusoidal long-period variation. It looks much more regular and of a stable character than $\omega_2(t)$, and has the same period as the accurately estimated $p_i=2737$ yr. 
Thus, the librational parts of secular perturbation for the outer planet have the same period for inclination $i$, periastron argument $\omega$, and nodal longitude $\Omega$.
In stark contradiction to the general predictions of the quadrupole approximation, there is no discernible nodal recession of the outer planet in this configuration.

\begin{figure}[H]

\begin{adjustwidth}{-\extralength}{0cm}
\centering 

\includegraphics[width=.62\textwidth]{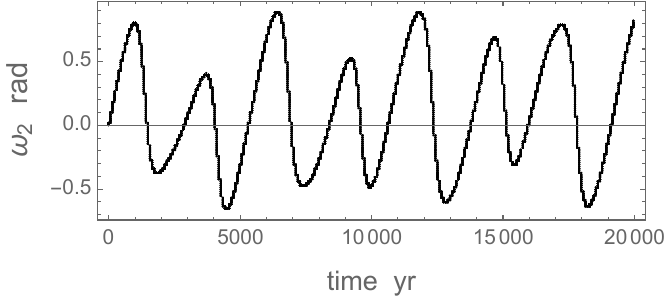}
\includegraphics[width=.62\textwidth]{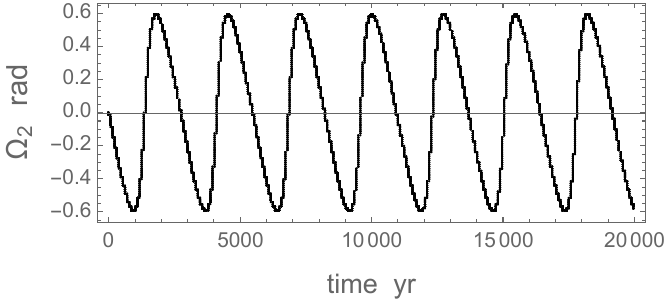}
\end{adjustwidth}
\caption{{Temporal} 
 behavior of K2-18 planets' nodes and apsides in a two-planet configuration. The numerical integration of orbital motion was performed for 30 Kyr with
the initial conditions: \mbox{$e_1(0)=0.01$,} $e_2(0)=0.05$, $i_1(0)=0$, $i_2(0)=5^\circ$, $\omega_1(0)=\omega_2(0)=0$, and $\Omega_1(0)=\Omega_2(0)=0$. 
\textbf{Left} panel: outer planet K2-18 b periastron argument after subtraction of a constant-rate precession of $0.0277^\circ$/yr. \textbf{Right} panel: outer planet K2-18 b longitude of the ascending node. Both angles are in~radians. \label{omegas.fig}}
\end{figure}
\vspace{-6pt}
\section{Transit Time Variations in a Two-Planet K2-18 System}
The chronology of planetary transits in front of the stellar disk gives us a rare possibility to observe the effects of orbital dynamics. K2-18 b is not known to show any variations in transit times, which are generally more difficult to detect in the extended mission data than in the main Kepler mission data. We approach this problem from the theoretical side, trying to predict the magnitude of possible variations and their time scale. In a two-planet system with the proposed inner companion (planet c, or planet 1 in this paper), the gravitational interaction between the planets results in the variability of all orbital elements (Section \ref{ex.sec}). Our high-accuracy numerical simulations of the orbital motion confirm that the semimajor axis ($a$) and, therefore, the mean motion ($n$), change in a very narrow interval with time. The exchange of orbital momentum between the planets is mostly seen in the periodic variation of eccentricity and inclination, and the more complex behavior of the geometric orientation angles $\Omega$ and $\omega$. The former element is the angle between a fixed (inertial) direction of the $x'$ Cartesian axis and the ascending node of the orbit. The periapse argument $\omega$ is the angle in the instantaneous plane of the orbit between the ascending node and the periapse of the planet. In the classical Lagrange perturbation theory, the variations of each orbital element are assumed to be separable into three independent categories: short-period variations, long-period variations, and secular drift. To approximate the latter two, the complex perturbation equations are averaged over one orbital period. The secular drift is assumed to be linear in time. Our numerical results for a grid of initial configurations confirm that there is no secular drift in $n$, $e$, or $i$. The osculating angles $\Omega$ and $\omega$, on the contrary, show both long-period variations, secular drifts, and sudden jumps. To elucidate the nature of these changes, we consider the behavior of the angular orbital momentum vector for planet 2.

For small eccentricity, the magnitude of orbital momentum is
\eb 
h=\sqrt{G(M_s+M_2)a_2(1-e_2^2)}\simeq \sqrt{G(M_s+M_2)a_2}(1-e_2^2/2).
\ee 
{Therefore}
, in the first-order approximation, $h_2$ changes periodically with time in a similar manner to $e_2$ but with the opposite phase. The amplitude of this variation is relatively small, whereas the tilt variation of the vector $\boldsymbol h$ is much more substantial. Our fixed inertial reference frame is defined by the orbital plane of planet 1 at time $t=0$. The axes $x'$ and $y'$ of this frame are in this plane, and the third axis $z'$ completes the right-hand triad. Thus, axis $z'$ is close to the average direction of the planets' orbital momenta. The direction cosines of the vector, i.e., the projections of the normalized vector $\langle \boldsymbol{h}\rangle=\boldsymbol{h}/||\boldsymbol{h}\||$ are computed as
\begin{eqnarray}
    \langle h\rangle_x &=& \sin i_2 \sin \Omega_2 \nonumber \\
    \langle h\rangle_y &=& -\sin i_2 \cos \Omega_2 
\end{eqnarray}
from the osculating elements in the output file. The result for the initial configuration with  $e_1(0)=0.01$, $e_2(0)=0.05$, $i_1(0)=0$, $i_2(0)=5^\circ$, $\omega_1(0)=\omega_2(0)=0$, and $\Omega_1(0)=\Omega_2(0)=0$ is shown as a parametric curve for each planet in Figure~\ref{hh.fig}. Both orbits describe complicated folded loops
in this projection. The inner planet 1 (red curve) acquires significantly greater orbital inclinations than the outer planet. The picture is strikingly different from a regular nodal recession, which has been assumed in many analyses. A secular recession of the nodes with a constant inclination would show
as a concentric circle on this plot. We note that the trajectory of the inner planet crosses the origin several times during the simulated interval of 30 Kyr. In these instances, the inclination of planet 1~nullifies, and the osculating $\omega$ and $\Omega$ jump by $\pi$. This could alternatively be represented by continuous functions allowing the inclination to acquire negative values (which formally violates the definition of the ascending node). Keeping this technical feature in mind, we can now compute the
observable transit time variations (TTV) of planet 2.

The commonly used formula for consecutive transit time $t_k$ from \citep{1995Ap&SS.226...99G}  in the presence of periastron precession is 
 \eb
 t_k\simeq t_0+P_{\mathrm{orb}}\,k-\frac{e\,P_{\mathrm{orb}}}{\pi}\cos\left(\omega_0+\frac{d\omega}{dk} k\right)+O(e^2),
 \ee
 where $\omega_0$ is the argument of the periastron at time $t_0$, and $k$ is the scaled time (orbit counter) equal to $t/P_\mathrm{orb}$. This formula is not valid in our case.
 First, we note that it is derived as a first-order approximation specifically for the case $i=0$, and the $\omega$ appearing there is not the periastron argument but a compound variable $u=\Omega+\omega$, i.e., the true longitude of the periastron. It also implicitly assumes that the periastron precesses at a constant rate, while the nodal variation is nil. For a gravitationally perturbed planet, all the orbital elements are functions of time, and computing transit times are not straightforward. A transit time $T$ can be defined as the instance when the planet's trajectory projected on the fixed $\{x'\,y'\}$ plane, which includes both the vernal equinox and the line of sight, crosses the line of sight. Without a loss of generality, the $x'$ axis can be chosen as the line of sight, and the transit equation is then
 \eb
 x'(T)\equiv \boldsymbol{R}_3(-\Omega)\,\boldsymbol{R}_1(-i)\,\boldsymbol{R}_3(-\omega)\,[1,0,0]^T\,[[1]]=0,
 \ee
 where $\boldsymbol{R}_j(\alpha)$ denotes the right-handed Euler rotation matrix around axis $j$ through angle $\alpha$, and the time arguments $(T)$ of all the osculating elements are omitted for brevity. In the traditional trigonometric form, this equation is equivalent to
 \eb
 \begin{split}
 -\sqrt{1 - e^2} \sin{E} (\cos\Omega \sin\omega + \cos i \cos\omega \sin\Omega) \\
 + (\cos E-e) \cos\omega \cos\Omega - \cos i  \sin\omega \sin\Omega)=0,
 \label{treq.eq}
 \end{split}
\ee 
 where $E$ is the eccentric anomaly. Using the output states of the simulations files, which include the values $\Omega$, $\omega$, and $i$ on a regular cadence, the corresponding values $E(T)$ can be computed by finding the root of this equation by numerical methods. These values can be converted to the corresponding mean anomaly values by using Kepler's equation:
 \eb
 M=E-e\,\sin E.
 \ee
 {Finally}, the transit time $T$ is obtained from $M(T)$ and the corresponding times of the periastron passage $T_0$ (also given in the numerical integration output) by
 \eb 
 T=T_0+M/n.
 \ee 
  {Equation} (\ref{treq.eq}) has two roots for each recorded state, because the planet is aligned with the line of sight twice during each revolution, when it transits in front of the stellar disk, and when it is eclipsed behind the star. If the ascending node is defined as the point on the celestial plane where the planet moves toward the observer, the transit time corresponds to the root where $y'(T)$ is positive.
  
  \begin{figure}[H]

\includegraphics[width=.5\textwidth]{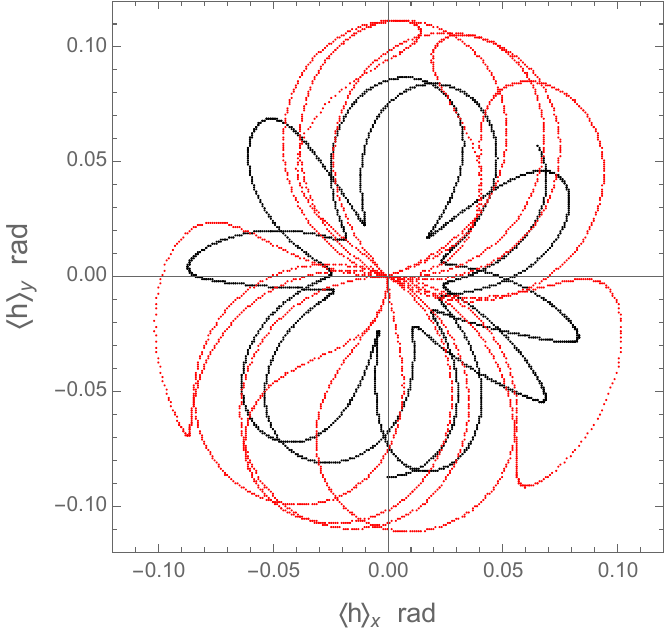}
\caption{{The} 
 paths of the orbital axes of K2-18 planets over 30 Kyr in the inertial $\{x',y'\}$ plane. The red curve is for the inner planet 1 (K2-18 c), 
and the black curve is for the outer planet 2 (K2-18 b). The same initial conditions are used for numerical integration as in
Figure~\ref{per.fig}.  \label{hh.fig}}
\end{figure}

 The resulting curve $T(t)$ is a wavy inclined line. It would be a perfectly straight line  for an unperturbed single planet. To see the effects of secular gravitational perturbations in the observable intervals between transits, which are called transit time variations (TTV), the linear trend (which is not observable) can be removed by taking the modulo function of the $T$ values:
 \eb 
 \Delta T={\rm Mod} [T, P'_{\rm orb}],
 \ee 
 where $P'_{\rm orb}$ is the empirically estimated average orbital period in the inertial frame (not to be confused with the often invoked anomalistic period). 
 Thus, the computed periodic part of the TTV for the initial configuration $e_1(0)=0.01$, $e_2(0)=0.05$, $i_1(0)=0$, $i_2(0)=5^\circ$, $\omega_1(0)=\omega_2(0)=0$, and $\Omega_1(0)=\Omega_2(0)=0$,
 which has also been used in Figures \ref{per.fig} and \ref{hh.fig}, is shown in Figure~\ref{ttv.fig} for the planet of interest (b). The empirically estimated $P'_{\rm orb}=32.940562$~d is slightly longer than the nominal (unperturbed) orbital period 32.94 d listed in Table~\ref{param.tab}. At first glance, the result is surprising because we see quasi-sinusoidal variations with a period of $\sim$4960 yr and a full amplitude of several days. The secular TTV period equals the period of eccentricity libration $P_e$, which is also the period of orbital momentum oscillation. This huge TTV variation, however, cannot be observed. The time span of currently available TTV data, mostly from the {\it Kepler} missions, is only a few to several years. The character of secular TTV oscillations is such that the planet spends most of the time on the ascending or descending segments of the curve, which are almost linear. A linear trend in transit times is not detectable, because it is automatically nullified by fitting a slightly adjusted $P'_{\rm orb}$.
 Only the epochs of the highest curvature are of interest, because they may be detectable. Still, for K2-18 c, the TTV measurements should span $>300$ yr for the curvature around the extrema of the transit time curve to become detectable at the current level of measurement precision (approximately, 20 min). 

\begin{figure}[H]

\includegraphics[width=.55\textwidth]{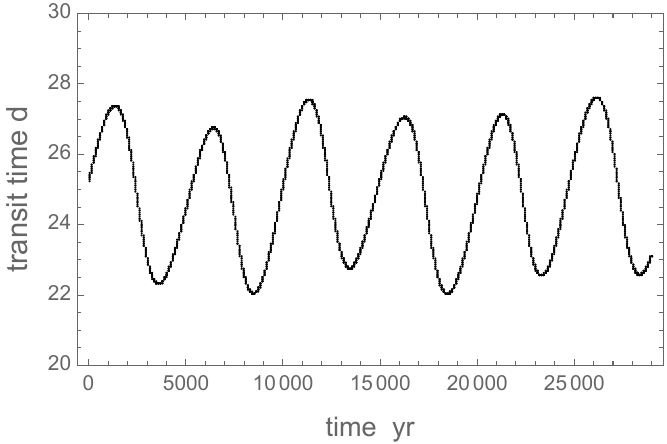}
\caption{{Transit} 
 time variations in days for planet K2-18 b computed from the same simulation of a two-planet system as shown in
Figure~\ref{per.fig}.  \label{ttv.fig}}
\end{figure}

In terms of measurable effects, which might be useful to confirm the presence of a perturbing inner planet, the short-period variations of transit times (and the orbital momentum magnitude) are more interesting. These are smaller by a few orders of magnitude than the LP oscillations, but their time scale is down to $\sim$10 yr. To elucidate the issue of TTV detectability, we performed a special integration for 30 Kyr with state dumps every year and the same initial parameters. The output files allow us to look closer at the character of short-term variations in transit times. We processed the simulated data in the same way as described above. The result for a small section of the curve spanning 30 yr is shown in Figure~\ref{1yr.fig}. The linear trend is not measurable, because it would be absorbed in the fitted average transit period. The remaining variations look non-periodic and have an amplitude up to 10 min. They should be present in the TTV measurement data, but it is not easy to recognize them because of the absence of a specific pattern. Most likely, such variations would be interpreted as excessive observational noise that is not captured by the formal~uncertainties. 

\begin{figure}[H]

\includegraphics[width=.55\textwidth]{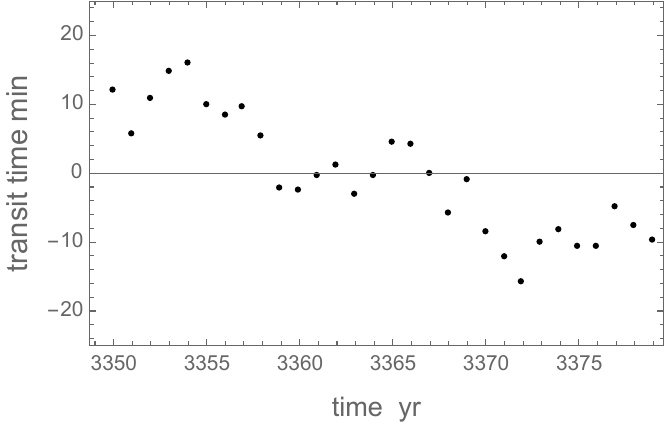}
\caption{{Transit} 
 time variations in minutes for planet K2-18 b computed from a special simulation of a two-planet system with the same initial parameters as in
Figure~\ref{per.fig}  for 30 Kyr with a state dump step of 1 yr. A small section of the data spanning 30 yr close to a minimum of the secular variation curve is shown.}
\label{1yr.fig} 
\end{figure}

\section{Tidal Orbital Damping in a Single-Planet K2-18 System}
\label{tide.sec}
If the inner planet c (planet 1 in this paper) is a false positive, and K2-18 is a single-planet system, the possible effects of the orbital perturbation scale down in magnitude and variety. A Keplerian orbit with fixed orbital elements is then a very good approximation, and the secular effects are caused by the finite size of the gravitating bodies and their rotation. The prevalence of low eccentricity orbits for detected exoplanets is in stark disparity with the statistics for binary stars, where a flat distribution of eccentricity is determined. It is widely assumed that the tidal dissipation mechanisms are responsible~\citep{2011MNRAS.414.1278P}. As the planet revolves around the star, the local gradient of the gravitational potential within the planet's body creates a transient distortion (asphericity) of its shape, which is called the tidal bulge. A stationary tidal bulge is realized when the perturber (the star in this case) is motionless with respect to the planet's surface. In a more realistic dynamical model, the bulge moves across the surface because of the combined effect of the perturber's orbital motion and the planet's rotation. The internal tidal friction is a nonlinear function of the relative angular velocity, and the resulting polar torque on the planet can be formally represented with an infinite series of Fourier harmonics in the more advanced Maxwell and Andrade rheological models \citep{1964RvGSP...2..661K, 2008CeMDA.101..171F, 2009CeMDA.104..257E}. As a mathematical abstraction, this can be viewed as a number of ``waves" running around the circumference with diminishing amplitudes and harmonics of the main tidal frequency. The physical reality is a single tidal deformation, which is mostly an ellipsoidal shape, with overlaid periodically-changing higher-order spherical harmonics. 

For simplicity, the perturber is considered to be a point mass, and the tidal torque acting on the planet is reciprocated by an orbital torque acting on the perturber. 
To complete the model, the tides raised by the planet on the star should be considered, in which case the planet is assumed to be a point mass. The total orbital action is
the sum of the torques generated by the planetary and stellar tidal bulges. The differential equations of orbital evolution are derived in the framework of the Lagrange
perturbation theory by averaging out the disturbing function over one orbital period. Thus, only the secular and long-period can be obtained in principle, but an additional averaging over one circulation period of the periastron leaves only the secular drifts. With these caveats and limitations, the resulting equations for the tides raised by the star on the planet are \citep{2019CeMDA.131...30B}:\vspace{12pt}
\begin{eqnarray}
\label{da.eq}
    \frac{da}{dt} & \simeq & -n\,\frac{M_s}{M_2} \frac{R_2^5}{a^4}\, \bigl[3 (1-5 e^2+\frac{63}{8} e^4) K_2(2 n-2 f_2) \\ \nonumber
    && +\frac{3}{8}e^2(1-\frac{1}{4}e^2)K_2(n-2 f_2)+\frac{9}{4}e^2(1+\frac{9}{4}e^2)K_2(n) \\ \nonumber
    && +\frac{81}{8}e^4 K_2(2n) +\frac{441}{8}e^2(1-\frac{123}{28}e^2)K_2(3n-2 f_2) \\ \nonumber
    && +\frac{867}{2}e^4 K_2(4n-2 f_2)\bigr], \\ 
    \frac{de}{dt} & \simeq & -n\,e\,\frac{M_s}{M_2} \frac{R_2^5}{a^5}\, \bigl[-\frac{3}{16} (1-\frac{1}{4} e^2) K_2(n-2 f_2) \\ \nonumber
    && -\frac{3}{4}(1-\frac{21}{4}e^2)K_2(2 n-2 f_2)+\frac{9}{8}(1+\frac{5}{4}e^2)K_2(n) \\ \nonumber
    && +\frac{81}{16}e^2 K_2(2n) +\frac{147}{16}(1-\frac{179}{28}e^2)K_2(3n-2 f_2) \\ \nonumber
    && +\frac{867}{8}e^2 K_2(4n-2 f_2)\bigr], 
\end{eqnarray}
where $f_2$ is the rotational frequency of planet 2, $R_2$ is its radius, $K_2$ is its frequency-dependent tidal quality function (kvalitet), $n$ is the orbital mean motion, $M_s$ is the mass of the star, and the other parameters are as previously defined. The contributions from the tides raised by the planet on the star can be obtained by swapping the subscripts 2 and $s$ everywhere. These are approximate equations that ignore all terms $O(\epsilon^2)$, $O(e^6)$, and higher orders of obliquity $\epsilon$ and eccentricity, also taking into account only the main (quadrupole) term of the perturbing potential. Even with these simplifying approximations, the resulting secular trend is quite uncertain in general, because the parameters $e$, $f$, and $K(\nu)$ are practically unknown. It is safe, however, to neglect the contribution from the star, because the kvalitet values for old M-dwarfs are expected to be very low. For gas giant planets, these values can be as high as $10^{-5}$--$10^{-4}$, and even larger for semi-molten terrestrial compositions
\citep{2018ApJ...857..142M}.

The existing models for $K_2(\nu)$ suggest complex functions with anti-symmetric kinks at the main spin--orbit resonances 1:1, 3:2, etc. In the better theoretically developed Maxwell rheological model, the amplitude of the kinks and their position depends on the so-called Maxwell time, which is the ratio of the effective viscosity and the unrelaxed shear  modulus~\citep{2012CeMDA.112..283E,2015ApJ...810...12M}. The driving parameter is the viscosity of the energy-dissipating layer of the planet. While this parameter is estimated to be in the range of $10^{-4}$--$10^{-2}$ Pa s for warm, pressurized water in the interiors of icy giants \citep{2019ApJ...881...81F}, pressurized ice has values of the order $10^{15}$ Pa s. The steep dependence of the effective viscosity on temperature for silicates also defines a great dynamical range of Maxwell times, and, therefore, of tidal qualities, for planets of a terrestrial composition. Planet K2-18 b is presumed to have an Earth-like core covered by a massive water ocean. The ocean is unlikely to be of significance in dissipating the tidal energy, and most of the action should be allocated to the terrestrial core. Given these prerequisites, we can consider three distinct options for the tidal spin--orbit interaction.

The first option is for a rapidly rotating planet's core of terrestrial rheology. If the solid body of K2-18 b rotates as fast as the Earth, $f_2>>n$, we can omit the multiples of $n$ in the tidal mode arguments of $K_2$, which obtains, also omitting the relatively small term proportional to $K_2(2 n)$:\vspace{12pt}
\begin{eqnarray}
\label{fast.eq}
    \frac{da}{dt}^{\rm (fast)} & \simeq & n\,\frac{M_s}{M_2} \frac{R_2^5}{a^4}\, \bigl[\frac{3}{8} (8+108 e^2+573 e^4) K_2(2 f_2) \\ \nonumber
    && -\frac{9}{4}e^2(1+\frac{9}{4}e^2)K_2(n) \bigr], \\ 
    \frac{de}{dt}^{\rm (fast)} & \simeq & n\,e\,\frac{M_s}{M_2} \frac{R_2^5}{a^5}\, \bigl[\frac{3}{4} (11+505 e^2) K_2(2 f_2) \\ \nonumber
    && -\frac{9}{8} (1+\frac{5}{4}e^2)K_2(n) \bigr].  
\end{eqnarray}
{Even} the sign of these derivatives is uncertain, because for sufficiently cold and inviscid planets, $K_2(n)>K_2(2f_2)$. The closest analogy of such systems is the Earth--Moon system, which is expanding both in $a$ and $e$. This is not, however, always the case, even for such cold and inviscid planets as Earth. When both frequencies $n$ and $2 f_2$ are far from the main resonances and the tidal quality peaks, $K_2(\nu)$ is approximately inversely proportional to frequency. In this case,
\eb 
K_2(n)/K_2(2 f_2)\approx 2 f_2/n.
\ee 
{A simple} calculation reveals that if the rotational period of K2-18 b equals 1 d, for example, $da/dt$ becomes negative for $e>0.165$. Orbital decay for such rapidly rotating planets is precluded for any eccentricity only if $f_2\lesssim 15\,n$. The relative contribution of the two terms in Equation (\ref{fast.eq}) is of the same order of maginitude, because they both are $O(e^0)$, and we find that the planet's orbit can circularize ($de/dt<0$) at small eccentricities if $f_2\gtrsim 4\,n$.

The Maxwell time and kvalitet values are quite uncertain for K2-18 b. We are using the $K_2(n)=
10^{-4}$ value as a benchmark, which is the upper envelope for astrometrically estimated values for Saturn and Jupiter. The computed derivatives are then, with the other parameters listed in Table~\ref{param.tab}, $(da/dt)^{\rm (fast)}\simeq 3\times 10^{-7}$ m d$^{-1}$ and $(de/dt)^{\rm (fast)}\simeq 0$ for $e=0$, and $(da/dt)^{\rm (fast)}\simeq 3\times 10^{-7}$ m d$^{-1}$ and $(de/dt)^{\rm (fast)}\simeq -3\times 10^{-18}$ m d$^{-1}$ for $e=0.1$. Thus, a secular expansion of the orbit and a slow circularization are possible at low eccentricity values, but the rates are marginally low, and it is therefore unlikely that the tidal torques play any role in the habitability conditions of the single K2-18 planet.

The second option considered in this paper is a 3:2 spin--orbit resonance. For planets of terrestrial compositions that are close to their host stars and have a finite orbital eccentricity, this is the more probable equilibrium state than the synchronous rotation (1:1 resonance). The closest example of a planet in a 3:2 spin--orbit resonance is Mercury \citep{1965Natur.206Q1240P}, although temporary captures into other resonances are possible over the course of Mercury's tumultuous dynamical history \citep{2014Icar..241...26N}. Higher chances of capture into the higher-order resonances are associated with larger orbital eccentricities, but the triaxial elongation of the permanent shape is also an important factor \citep{1966AJ.....71..425G, 2012ApJ...752...73M}. The differential equations of orbital evolution for the 3:2 resonance, without any additional omissions, become
\begin{eqnarray}
\label{3:2.eq}
    \frac{da}{dt}^{\rm (3:2)} & \simeq & n\,\frac{M_s}{M_2} \frac{R_2^5}{a^4}\, \bigl[(3-\frac{69}{4} e^2 - \frac{6639}{16} e^4) K_2(n) \\ \nonumber
    && +\frac{3}{32}e^2(4-109 e^2)K_2(2n) \bigr], \\ 
    \frac{de}{dt}^{\rm (3:2)} & \simeq & -n\,e\,\frac{M_s}{M_2} \frac{R_2^5}{a^5}\, \bigl[\frac{3}{32} (20+1129 e^2) K_2(n) \\ \nonumber
    && +\frac{3}{64} (4+107 e^2)K_2(2n) \bigr].  
\end{eqnarray}
{The} orbit decays ($da/dt$ is negative) for $e\gtrsim 0.26$ and expands for a smaller eccentricity. The orbital eccentricity always decreases, irrespective of its current value. The estimated rates for a cold terrestrial planet with $K_2(n)=10^{-4}$ are too low to be of consequence. We estimate 
$(da/dt)^{\rm (3:2)}\simeq 4.5\times 10^{-6}$ m d$^{-1}$ and $(de/dt)^{\rm (3:2)}\simeq -2.3\times 10^{-17}$ d$^{-1}$ for $e=0.1$, and $(da/dt)^{\rm (3:2)}\simeq 2.7\times 10^{-6}$ m d$^{-1}$ and $(de/dt)^{\rm (3:2)}\simeq -9.6\times 10^{-17}$ m d$^{-1}$ for $e=0.2$.

Finally, the third option is a synchronous rotation, i.e., the 1:1 spin--orbit resonance. This equilibrium state is often assumed in the literature for habitable exoplanets. However, this equilibrium state corresponding to the global minimum of potential energy should not be taken for granted for planets of terrestrial composition with significant triaxiality of the inertia tensors \citep{1966AJ.....71..425G}. In particular, it has been shown that higher-order spin--orbit resonances are more probable for specific two-planet systems with habitable planets \citep{2014ApJ...780..124M}. Still, for the sake of consistency, we provide the approximate equations for this case:
\begin{eqnarray}
\label{1:1.eq}
    \frac{da}{dt}^{\rm (1:1)} & \simeq & n\,\frac{M_s}{M_2} \frac{R_2^5}{a^4}\, \bigl[3 e^2 (-19+79 e^2) K_2(n) \\ \nonumber
    && -\frac{3549}{8}e^4K_2(2n) \bigr], \\ 
    \frac{de}{dt}^{\rm (1:1)} & \simeq & n\,e\,\frac{M_s}{M_2} \frac{R_2^5}{a^5}\, \bigl[\frac{3}{8} (-28+153 e^2) K_2(n) \\ \nonumber
    && -\frac{1815}{16}  e^2 K_2(2n) \bigr].  
\end{eqnarray}
{Note} that the largest terms for $da^{\rm (1:1)}/dt$ are $O(e^2)$, which almost nullifies the orbit size evolution at a small or moderate eccentricity. The sign of $de^{\rm (1:1)}/dt$ changes from minus to plus at $e\approx 0.4$, but our formulae become inaccurate at such eccentricity values, and the 1:1 synchronization is hardly possible for terrestrial planets anyway. The estimated rates of eccentricity decay are still vanishingly low, amounting to $\sim$10$^{-16}$  d$^{-1}$.

More complicated scenarios of tidal evolution are possible if planet K2-18 b is covered with a deep global water ocean. The liquid outer layer does not have a permanent figure, and it is likely to be rotating pseudosynchronously \citep{2015ApJ...810...12M}, if it is mechanically decoupled from the rocky bottom. The rocky body, on the other hand, is either synchronized or rotates faster than the ocean. Depending on the coefficient of friction between the ocean and the rocky core, a significant additional mechanism of energy dissipation can emerge. Similar models, but with a reversed structure, including a molten core and a solid outer shell, have been investigated for Mercury \citep{2007Sci...316..710M}.

Our conclusion is that, irrespective of the rotational state of the planet, tidal dissipation can hardly be significant in shaping the orbit in the long term. If the eccentricity is indeed very small, there should be other reasons for that. The dynamically cool population orbiting M-dwarfs with small eccentricities \citep{2023PNAS..12017398S} can be primordial, i.e., inherited from the early stages of migration in the protoplanetary disk. This result may change if the inner planet in K2-18 is real. Because of the explicit $n(R/a)^5\propto R^5/a^{13/2}$ dependence of secular rates in \mbox{Equation (\ref{da.eq})}, the rates can be higher by 2--3 orders of magnitude for the inner companion. This makes the inner planets relatively efficient sinks of the orbital energy and angular momentum. The eccentricity exchange mechanism (Section \ref{ex.sec}) prevents circularization of the inner planet as long as the outer planet remains eccentric. Therefore, the tidal damping of the inner planet is effectively transferred to the outer planet as well. This makes the characteristic $e$-folding times somewhat longer, but they may be comparable with the lifetimes of M-dwarfs.

\section{Secular Perturbations Caused by Stellar Rotation}
\label{rot.sec}
The equilibrium shape of a rotating star is an axially symmetric oblate figure. The gravitational potential outside the stellar surface is formally represented by an expansion in spherical harmonics, where the lowest-order zonal (quadrupole) term is by far the greatest. The disturbing function generated by this deviation from spherical symmetry is proportional to the corresponding coefficient $J_2$, which is the quadrupole moment. Ignoring the small difference between the total mass $M_s+M_2$ and $M_s$, the equation is
\eb 
J_2=\frac{k_s}{3}\left(\frac{f_s}{n}\right)^2 \left(\frac{R_s}{a}\right)^3,
\ee 
where $k_s$ is the corresponding Love number and $f_s$ is the frequency of the stellar rotation. Using the nominal values in Table~\ref{param.tab} and assuming $P_{\rm rot}=8.8$ d,  $J_2\approx 1.6\times 10^{-5}\,k_s$ is estimated for the star K2-18. The Love number is quite uncertain, but we can assume it to be of the order of 0.1, which would bring $J_2$ to approximately $1.6\times 10^{-6}$, which is almost eight times the time-average $J_2$ value of the Sun \citep{2023ApJ...942...90E}. 

The expansions for the disturbing functions and Lagrange planetary equations have been worked out up to $O(e^4)$ and the $J_6$ zonal harmonics for the significantly more perturbed geopotential \citep{1961mcm..book.....B, 1961GeoJ....4...17M}. As usual, the orbital effects are supposed to be separable into long-period and secular terms. For the latter, limiting our consideration only to the leading terms, the equations are:
\begin{eqnarray}
\label{sec.eq}
    \frac{d\Omega}{dt}^{\rm (sec)} & \simeq & -\frac{3 n}{2 p^2}\,\cos \epsilon\, J_2 \\ \nonumber
    \frac{d\omega}{dt}^{\rm (sec)} & \simeq & \frac{3 n}{p^2}\,\bigl(1-\frac{5}{4} \sin^2\epsilon\bigr) J_2  \\ \nonumber
    \frac{d M}{dt}^{\rm (sec)} & \simeq & n+\frac{3 n}{2 p^2}\,\sqrt{1-e^2} \bigl(1-\frac{3}{2} \sin^2\epsilon\bigr) \,J_2,
\end{eqnarray}
where $\epsilon$ is the obliquity of the planet's orbit on the equator of the star, and $p=(1-e^2)\,a/R_s$. Note that at obliquity values below $63.43^\circ$, the sign of $(d\omega/dt)^{\rm (sec)}$ is positive, which signifies a constant-rate precession of the line of apsides. The sign of $(d\Omega/dt)^{\rm (sec)}$, however, is always negative. Thus, stellar rotation causes a nodal recession. The perturbation of the mean anomaly rate is a constant offset reflecting the additional gravitational acceleration of the planet in the radial direction, which is positive (faster mean motion) for $\epsilon< 54.74^\circ$. As we will see in the following, the equation for $(d\omega/dt)^{\rm (sec)}$ is of special importance, which, after substituting the formula for $J_2$, becomes
\eb 
\frac{d\omega}{dt}^{\rm (sec)} \simeq \frac{k_s n}{(1-e^2)^2}\,
\left(\frac{f_s}{n}\right)^2 \left(\frac{R_s}{a}\right)^5 (1-\frac{5}{4} \sin^2\epsilon).
\label{dom.eq}
\ee
{This} equation differs by a factor of two from the analogous formulae given in \citep{1939MNRAS..99..451S, 2009ApJ...698.1778R}. In these papers, the marginal case of $\epsilon=0$ is considered, and the quoted equations are given not for the periastron argument $\omega$ but for the compound variable $u=\Omega+\omega$. This subtle detail is crucially important for the general consideration with $\epsilon\neq 0$. There are no secular drifts of other orbital elements.

Long-period variations emerge for all the orbital elements. The approximate equations are obtained by the integration of periodic terms of the corresponding derivatives over time. All these relations are proportional to $e\,J_2$ except the two most important elements in the context of habitability conditions, eccentricity, and the periastron argument. Limiting the expansion to the dominant (for small $e$) main harmonics in $\omega$, the appropriate equations~are:
\begin{eqnarray}
\label{LP.eq}
    \Delta\omega^{\rm (LP)}  & \simeq & \frac{J_2}{8\,e\, p^2}\,\bigl((-12-25 e^2)+8(1+4 e^2) I \bigr)\sin \omega  \\ \nonumber
    \Delta e^{\rm (LP)} & \simeq & \frac{J_2}{8 p^2}\,\bigl(3(4+e^2)-8(1-e^2) I \bigr)\cos \omega,
\end{eqnarray}
where $I=\sin^2\epsilon$. In the right-hand parts of these equations, $\omega$ includes only the secular motion term. Note that the amplitude of $\omega$ oscillations can be much larger than the constant-rate precession over one precession period for $e<0.5$. At a very small $e$ and $\epsilon$, the LP variation becomes the dominant term. Its amplitude can be arbitrarily large for $e\rightarrow 0$. This, however, does not make the LP perturbation more detectable, because the observable TTV effect is proportional to $e$. 

Assuming $J_2=1.6\times 10^{-6}$ for the star K2-18, we estimate the period of periapse precession driven by the rotational deformation to be approximately equal to 84 Myr. Such small effects cannot be measured today. The amplitude of LP oscillation of $\omega$ is a meager $5.4\times 10^{-7}$ rad for $e_2=0.001$, and $1.1\times 10^{-7}$ rad for $e_2=0.005$. The amplitude of the eccentricity oscillations is also negligibly small. We conclude that the orbit of planet K2-18~b should be extremely stable and nearly constant, if it is the only planet in the system.

\section{Milankovi{\'c} Cycles of Habitable Exoplanets}
Gravitational and tidal interactions of Earth with the Sun, the Moon, and other planets produce cyclic and secular perturbations of the orbit and the rotation axis, which modulate the climate ({\url{https://climate.nasa.gov/news/2948/milankovitch-orbital-cycles-and-their-role-in-earths-climate/}, {accessed on 6 October 2023}). The fastest cycle with a period of $\sim$26 Kyr is related to the nodal precession caused by the gravitational pull of the Sun and the Moon on the oblate shape of Earth with a significant $J_2$ coefficient. The climate impact is limited to the time of the perihelion passage in the year. It combines with the much slower apsidal precession $d\omega/dt^{\rm (sec)}$ to produce a constant drift of the seasons within the year with a period of $\sim$21~Kyr, as well as the modulation of the relative duration of the seasons. The cold and warm seasons are equal in duration when $\omega=0$ or $\pi$. Perhaps of greater importance are the secular changes in the obliquity of the Earth's equator on the ecliptic in the range $2.4^\circ$. A larger obliquity causes greater seasonal variations of insolation and temperature. The period of this oscillation is $\sim$41 Kyr. Finally, as the Earth's orbit interacts with the other planets' orbits, the true inclination with respect to the invariable plane oscillates with a period of approximately 100 Kyr. The climate impact is similar to the long-period variation of the obliquity. Although this effect is believed to be relatively small, the timescale is conspicuously close to the main cycle of glaciation periods \citep{1976Sci...194.1121H}. The original Milankovi{\'c} model includes the oscillations of orbital eccentricity (of similar periods), and indeed, both numerical integrations and the Laplace--Lagrange theory show that Earth's eccentricity varies between 0 and 0.06 and its inclination varies between 0~and $2.5^\circ$ in a quasi-cyclic manner with characteristic periods close to 100 Kyr  \citep{2021A&A...655A...1M}. The shape of each cycle is remarkably similar to the curves in our Figure~\ref{per.fig} for the inner planet K2-18 c, indicating that the same basic dynamics are at work.  Empirically, these relatively small variations of orbital inclination and eccentricity emerge as the most important factors driving the ice ages on Earth.

Similar processes take place in systems with two or more exoplanets \citep{Ershkov2023}. As far as the system K2-18 is concerned, the options with Milankovi{\'c}-like cycles are quite different for the single-planet and two-planet configurations. If planet b is a lone planet, both the tidal perturbation of the planet and the secular perturbations caused by the rotation of the star are too slow and too small to be of significance (Sections \ref{tide.sec} and \ref{rot.sec}). The remaining important parameters, which are also the most uncertain, are the rate of the rotation of the planet and its obliquity on the orbital plane. The planet may be covered by a deep global water ocean, which is most likely to rotate pseudosynchronously, i.e., slightly faster than the mean orbital motion. The rocky core may be dynamically decoupled from the ocean, and its rotation may initially be different, including a finite obliquity. The force of friction at the bottom of the ocean is probably sufficient to equalize this differential rotation over the age of the system. However, the core has a permanent figure with triaxial elongation, which precludes the pseudosynchronous equilibrium. It can remain captured in a 3:2 or 1:1 spin--orbit synchronicity. The emerging picture is a constant circulation of the global ocean superimposed with periodic flows caused by the forced librations of the solid core. In the context of climate changes, a water-world planet is unlikely to have seasons, and it would probably have permanent surface ice covering the polar caps, but the insolation flux and temperature is more uniformly distributed around the equatorial zone by the slow residual rotation of the ocean with respect to the direction to the star.

Possible scenarios of Milankovi{\'c}-like cycles in a two-planet K2-18 systems are rich and diverse. Apart from the secular nodal recession and apsidal precession, relatively fast periodic variations of the habitable planet b are predicted in eccentricity and inclination. Their characteristic periods are a few thousand years. The tidal damping of eccentricity is somewhat more efficient because of the constant eccentricity exchange between the planets, and the much higher dissipation rate for the inner planet 1. The constantly changing inclination precludes an equilibrium state with zero obliquity, which results in rapidly changing seasons, smaller ice caps around the poles, and a climate that is more temperate and amenable to life. If the amplitude of the eccentricity cycles is significant, the outer planet's ocean will be warmed up and cooled down every 5000 yr. Furthermore, since the rate of pseudosynchronous rotation is defined by eccentricity, the long-term oscillation of the global ocean rotation and the current patterns should be expected.

\section{Summary}
\label{sum.sec}
We have investigated the dynamical evolution of the K2-18 exoplanet system using numerical and analytical methods in two possible configurations: a single planet b (detected via transits in front of the stellar disk), and a two-planet system including the suggested RV-detected planet c. Our main conclusions are:
\begin{enumerate}
    \item In the two-planet option, a relatively rapid exchange of angular momentum, eccentricity, and inclination between the planet is found with a characteristic cycle period of a few Kyr. The integrated trajectories are in excellent agreement with the first-order Lagrange perturbation theory developed for the Solar system planets.
    \item
    In the two-planet option, the lines of apsides of both planets are subject to both precession and long-period variations. The periods of the periodic components are non-commensurate. The line of nodes of the outer (habitable) planet undergoes long-period oscillations, but no significant secular drift. 
    \item 
    In the two-planet option, the vectors of orbital momenta describe complicated trajectories in the inertial frame because of the simultaneously varying inclination and nodes. The resulting secular TTV changes are large in amplitude (several days peak-to-peak) but so slow that they cannot be detected. The short-term TTV effects on the time scale of 1 year are marginally detectable, but they lack a specific pattern.
    \item 
    In the single planet option, the tidal damping of the eccentricity and orbit size is found to be negligibly slow for three possible rotational states: viz., a synchronous rotation---a 3:2 spin--orbit resonance---and a fast or retrograde rotation. The estimated characteristic $e$-folding times are longer than the stellar age.
    \item 
    The main $J_2$ moment of the star K2-18 due to its sidereal rotation is estimated at \mbox{$\sim1.6\times 10^{-6}$.} It turns out to be too small to excite a secular nodal recession or periastron precession that could be measurable. The amplitude of long-period variations is very small too.
    \item Milancovi{\'c}-like variations of planet b's climate are not expected in the single planet mode, but habitability conditions can be harsh, unless a significant obliquity is maintained. A rich variety of climatic changes emerges in the two-planet option with a rapid succession of seasons, constant circulation of the global ocean, a more temperate equatorial zone, and a long-period partial glaciation.
\end{enumerate}

\vspace{6pt}

\authorcontributions{{Conceptualization, V.V.M. and A.G.; methodology, V.V.M.; software, A.G.; validation, V.V.M.; formal analysis, V.V.M. and A.G.; investigation, V.V.M. and A.G.; resources, A.G.; writing---original draft preparation, V.V.M.; writing---review and editing, V.V.M. and A.G.; visualization, V.V.M.; supervision, V.V.M. All authors have read and agreed to the published version of the manuscript.} 
}

\funding{{This research received no external funding.} 
}

\dataavailability{The simulation inputs and results discussed in this paper are available upon reasonable request to the corresponding {author.} 
}

\conflictsofinterest{{Author Alexey Goldin was employed by the company Teza Technology. The remaining author declares that the research was conducted in the absence of any commercial or financial relationships that could be construed as a potential conflict of interest.} 
} 

\begin{adjustwidth}{-\extralength}{0cm}

\reftitle{References}

\PublishersNote{}
\end{adjustwidth}

\label{lastpage}
\end{document}